\documentclass[12pt,a4paper]{article}
\usepackage{amssymb}
\usepackage{graphicx}
\usepackage{amsmath}
\usepackage{placeins}

\begin{document}

\title{Scaling laws for the tropical cyclone derived from the stationary
atmospheric vortex equation}
\author{Florin Spineanu and Madalina Vlad \\
Association Euratom-MEC Romania, NILPRP \\
MG-36, Magurele, Bucharest, Romania \\
E-mail: spineanu@nipne.ro}
\maketitle

\begin{abstract}
In a two-dimensional model of the planetary atmosphere 
the compressible convective flow of vorticity represents a strong nonlinearity
able to drive the fluid toward a quasi-coherent vortical pattern. This is similar
to the highly organised motion generated at relaxation in ideal Euler fluids. 
The problem of the atmosphere is however fundamentally different since now there is an
intrinsic length, the Rossby radius. Within the Charney Hasegawa Mima model it has been
derived a differential equation governing the stationary, two-dimensional, highly organised
vortical flows in the planetary atmosphere. 
We present results of a numerical study of this differential equation. The most
characteristic solution 
shows a strong similarity with the morphology of a tropical cyclone.
Quantitative comparisons are also favorable and several relationships can be
derived connecting the characteristic physical parameters of the
tropical cyclone: the radius of the eye-wall, the maximum azimuthal velocity and the
radial extension of the vortex.
\end{abstract}

\tableofcontents



\section{Introduction}

In the absence of
dissipation the two-dimensional models of the
planetary atmosphere and of the magnetized plasma  
can be reduced to differential equations having the
same structure: the Charney equation for the nonlinear Rossby waves, in the
physics of the atmosphere (Charney 1948); and the Hasegawa-Mima equation for
drift wave turbulence, in plasma physics (Hesagawa and Mima 1978). One of
the characteristics of these equations is the presence of an intrinsic
length parameter, which can be seen from the fact that there is no space
scale invariance, in contrast to the Euler equation (see Horton and Hasegawa
1994). This length is the Rossby radius for the atmosphere and the ion sonic
Larmor radius for the magnetized plasma.

A long series of observations, experiments and theoretical (analytical and
numerical) studies has established that the fluids exhibit an intrinsic
trend to organisation. This is most obvious at relaxation from turbulent
states when the fluid evolves toward a reduced subset of flow patterns,
characterized by regular form of the streamfunction as shown for example by
Mathaeus \emph{et al.} 1991a, Matthaeus et al. 1991b, Kinney \emph{et al}.
1995, Horton and Hasegawa 1994 (and references therein).

It has been shown (Spineanu and Vlad 2005) that the stationary states
attained at late times in the evolution of the Charney-Hasegawa-Mima (CHM)
fluids are described by the following equation 
\begin{equation}
\Delta \psi +\frac{1}{2p^{2}}\sinh \psi \left( \cosh \psi -p\right) =0
\label{eq}
\end{equation}
Here $\psi $ is the streamfunction and $p$ is a positive constant. This
equation has been derived within a field-theoretical formulation of the
model of interacting point-like vortices and reveals that the asymptotic
state of organization of the physical vorticity in fluids and plasmas is
identical to states of self-duality of classical field theory of matter
interacting with a gauge field. Consistent with the physical model from
which it is derived, this equation should describe the states of
fluids/plasmas characterized by the presence of a background of vorticity (a
condensate of vorticity) and a finite intrinsic length (or a velocity of
propagation of perturbation, like the gravity wave speed or the sonic
speed). The numerical solutions of this differential equation has provided
interesting results for the physical systems which it should be able to
describe: plasma vortices, atmospheric vortices, non-neutral plasma vortex
crystals, current sheets. Comparisons with experimental results on
Navier-Stokes fluid, consisting of scatterplots $\left( \psi ,\omega \right) 
$ $\equiv $ (streamfunction, vorticity) (de Rooij et al. 1999), and with the
scatterplots obtained in numerical simulations by Seyler (1995) are also
favorable. A brief discussion of the derivation of Eq.(\ref{eq}) will be
presented in Section 2.

\bigskip

Due to the similar analytical structure (although for largely different
magnitude of parameters) of the plasma vortex and the atmospheric vortex, we
expect that this equation leads to solutions that may capture the strong
nonlinear character of the atmospheric vortical flows in those states where
the stationarity can be assumed a good approximation. Certainly the problem
of the structure of the atmospheric vortex cannot be reduced in general to
only fluid nonlinear dynamics, knowing the very important role of the heat
exchange and moisture transport and condensation processes. These are
essential elements of tropical cyclogenesis (see Emanuel 1986 and 1989 and
references therein) but it is often accepted that the fluid dynamics is an
adequate framework to study the vortex structure at the late stage of the
evolution (Reasor and Montgomery 2001, Kossin and Schubert 2001). The fluid
dynamic nonlinearity becomes the dominant constraint determining the
structure of the velocity field when the thermodynamic processes have
reached the stationary equilibrium.

The results of a numerical investigation of equation (\ref{eq}), for a range of parameters
relevant for the physics of the atmosphere, are summarised here.

The monopolar solutions of the differential equation have the same
morphology as the two-dimensional flow of a tropical cyclone\ (Willoughby
and Black 1996, Wang and Wu 2004, Reasor and Montgomery 2001, Kossin and
Schubert 2001). The solutions are characterized by a very narrow dip of the
profile of the azimuthal velocity (tangential wind) in the center of the
vortex. The radius of the ``maximum tangential wind'' or the radius of the 
\emph{eye wall} is much smaller than the radius of the vortex. There is a
decay of the magnitude of the azimuthal velocity toward the periphery, which
is much slower compared with the fast decay toward the center. We find a
very low magnitude (almost vanishing) of the vorticity over most of the
vortex (approx. from the radius of maximum wind to the periphery), while the
magnitude in a narrow central region is extremely high. Quantitatively, we
obtain for the diameter of the cyclone's eye a magnitude which compares well
with the observations. The maximum vorticity is in a realistic range and the
radial profile of the tangential velocity is similar to what is found in
observations or with what is obtained in empirical models and numerical
simulations. These favorable comparisons are valid in the case of mature,
quasi-stationary tropical cyclones, after the phases of genesis and dynamic
intensification.

Besides the structure of the flow, the numerical solutions for a very large
number of cases provide a large data set from which various correlations can
be extracted. In this regard we are mainly led to look for correlations
between parameters that may present interest in practical cases and
eventually can be compared with the observations. The main parameters that
have been collected in the numerical work are: total radial extension of the
atmospheric vortex, $R_{\max }$; the radius where the maximum of the
tangential velocity is attained, $r_{v_{\theta }^{\max }}$; the magnitude of
the maximum tangential wind, $v_{\theta }^{\max }$. In addition we have
collected the total energy $E_{fin}$ and vorticity $\Omega _{fin}$ in the
final state flow field. It is encouraging that we find the linear
relationship, $E_{fin}\sim \Omega _{fin}$ which has also been revealed in
numerical investigations of a large ensemble of point-like vortices
(Yatsuyanagi \emph{et al.} 2005). The numerical studies carried out until
now are organized in the form of several scaling relations connecting vortex
characteristics to the few parameters of the Eq.(\ref{eq}). They may be used
for comparison with observation or with more complex theoretical models.

In addition, the numerical study of Eq.(\ref{eq}) reveals the existence of
metastable states (quasi-solutions) consisting of (a) extremely concentrated
vortical flows, similar to the cross section of a tornado and (b) collection
of vortices with symmetric positions in plane (vortex crystals).

\section{The physical model and the field theoretical description}

In this section we briefly recall the origin of Eq.(\ref{eq}). It is useful
to consider first the ideal fluid described by the Euler equation, for which
it has been shown that it evolves at relaxation toward a very ordered flow
pattern, consisting of two (positive and negative) vortices. This state
persists for long times, being limitted by only the effect of some residual
dissipation. From numerical simulations it has also been inferred the form
of the flow function. It has been found that the streamfunction $\psi \left(
x,y\right) $ obeys, in these states, the \emph{sinh}-Poisson equation, an
equation which has very special properties. It is an exactly integrable
equation (by Inverse Scattering Transform) and is connected with a wide
class of fundamental systems, like the Thirring lattice of spins in plane,
the affine Toda system, the Gauss-Codazzi equations for embedding a surface
in three dimensional space, etc. David Montgomery and his collaborators have
developed a theoretical statistical model which explains the presence of
this equation in this context (Kraichnan and Montgomery 1980, Fyfe,
Montgomery and Joyce 1976, Joyce and Montgomery 1973, Montgomery and Joyce
1974, Montgomery \emph{et al}. 1992).

The Euler equation is 
\begin{equation}
\left[ \frac{\partial }{\partial t}+\left( -\mathbf{\nabla }_{\perp }\psi
\times \widehat{\mathbf{n}}\right) \cdot \mathbf{\nabla }_{\perp }\right] 
\mathbf{\omega }=0  \label{Euler}
\end{equation}
where $\widehat{\mathbf{n}}$ is the versor perpendicular on the plane, the
velocity and the vorticity are respectively $\mathbf{v}=-\mathbf{\nabla }%
_{\perp }\psi \times \widehat{\mathbf{n}}$ and $\mathbf{\omega }=\widehat{%
\mathbf{n}}\mathbf{\nabla }_{\perp }^{2}\psi $. It is generally accepted
(but not yet mathematically proved) that the Euler fluid may be described by
an equivalent model, consisting of a set of discrete point-like vortices
moving in plane under the effect of mutual interaction. The latter is
expressed by a potential given by the natural logarithm of the relative
distance between vortices normalized to the linear extension of the region
in plane where the motion is bounded (Kraichnan and Montgomery 1980). A
fundamental observation is that this formulation exhibits a particular
structure: matter, field, interaction. The \emph{matter} corresponds to the
density of point-like vortices in plane; the \emph{field} is derived from
the potantial and can be seen as an independent component of the system; the 
\emph{interaction} appears as the usual combination of the matter current
with the potential and leads to the equation of motion of the point-like
vortices. The \emph{sinh}-Poisson equation has been derived by formulating
the continuum version of the point-like vortices model as a field
theoretical model of interacting gauge and matter fields in the adjoint
representation of $SU\left( 2\right) $ (Spineanu and Vlad 2003). The
essential point of the latter derivation was the self-duality of the
relaxation states of the fluid.

In its simplest form the Charney-Hasegawa-Mima equation is 
\begin{equation}
\frac{\partial }{\partial t}\left( 1-\mathbf{\nabla }_{\perp }^{2}\right)
\psi -\left[ \left( -\mathbf{\nabla }_{\perp }\psi \times \widehat{\mathbf{n}%
}\right) \cdot \mathbf{\nabla }_{\perp }\right] \mathbf{\nabla }_{\perp
}^{2}\psi =0  \label{CHM}
\end{equation}
Similar to the Euler equation there is a discrete vortex model for the
Charney-Hasegawa-Mima equation, where the potential of interaction is now
short range (the modified Bessel function of zero order), \emph{i.e.} it
decays on the range given by the intrinsic length of the CHM equation. This
model has been used in meteorology by Morikawa (Morikawa 1960) and Stewart
(Stewart 1943).

In a similar approach as for the Euler fluid, it has been developed
(Spineanu and Vlad 2005) a field theoretical model for the continuous
version of the point-like vortices with short range interaction, based on
the Chern-Simons action for the gauge field in interaction with the
nonlinear matter field, in the adjoint representation of the $SU\left(
2\right) $ algebra. In this model it is possible to derive the energy as a
functional that becomes extremum on a subset of stationary states and
presents particular properties. The general characterization of this family
of states is their \emph{self-duality}, which here means that the energy
functional becomes minimum when the squared terms in its expression are all
vanishing, leaving as lower bound for energy a quantity with topological
meaning, proportional with the total vorticity.

The result is a set of equations parametrized by the solutions of the
Laplace equation in two-dimensions. The Eq.(\ref{eq}) is the simplest of
this family.

\section{Numerical studies}

To solve Eq.(\ref{eq}) we use the code ``GIANT A software package for the
numerical solution of very large systems of highly nonlinear systems''
written by U. Nowak and L. Weimann (Nowak and Weiman 1990). The code is part
of the numerical software library \emph{CodeLib} of the Konrad Zuse Zentrum
fur Informationstechnik Berlin. The meaning of the abbreviation is: GIANT =
Global Inexact Affine Invariant Newton Techniques. This code solves
nonlinear problems 
\begin{eqnarray}
F\left( \psi \right) &=&0  \label{gian} \\
\text{initial guess of solution, }\psi &=&\psi ^{\left( 0\right) }  \notag
\end{eqnarray}
where $F\left( \psi \right) $ is a nonlinear partial differential operator.
The presence of the hyperbolic trigonometric functions, with very fast
variation with the argument, renders the equation difficult to solve and
many initial conditions do not lead to a solution since they cannot initiate
a converging iteration. With all the difficulties of finding a right
initialization of the integration procedure we note however that the
solution with the morphology of a tropical cyclone appears insistently from
a wide class of initial functions which contains vortical flows.

\subsection{Parameters and Initializations}

Eq.(\ref{eq}) has been derived from the \emph{self-duality} subset of the
Euler-Lagrange variational equations for the action functional of the
field-theoretical model. In this action there are only two physical
parameters: the coefficient of the Chern-Simons action, which we have
identified as the sound speed, $c_{s}$ and the asymptotic vorticity which is
the Coriolis frequency $f_{0}$ (or, in plasma physics, the ion cyclotron
frequency $\Omega _{c}$). The space-like parameter that normalizes the
Laplace operator in the Eq.(\ref{eq}) is the ratio 
\begin{equation}
\rho _{g}=c_{s}/f_{0}  \label{rhog}
\end{equation}%
\emph{i.e.} the Rossby radius $\rho _{g}$ (respectively the sonic Larmor
radius in plasma, $\rho _{s}=c_{s}/\Omega _{ci}$). All distances implied in
the solution are normalized to $\rho _{g}$. The streamfunction is normalised
as 
\begin{equation}
\psi =\frac{\psi ^{phys}}{\rho _{g}^{2}f_{0}}  \label{psinorm}
\end{equation}%
The unit for the streamfunction is $\rho _{g}^{2}f_{0}$ and the unit for
vorticity $f_{0}$. Then the unit for velocity is $\rho _{g}f_{0}$. Here and
in the rest of the paper the upperscript \emph{phys} is used to indicate
that the quantity is dimensional, measured in appropriate physical units.
However the units themselves (\emph{i.e.} $\rho _{g}$, $f_{0}$ and
combinations) are written without this upperscript since there can be no
confusion.

If we know the large radial extension of a vortex ($R_{\max }^{phys}$) the
normalized parameter $R_{\max }$ is obtained by dividing to $\rho _{g}$. The
result of integration is very sensitive to $R_{\max }$ and this points out
the essential role played by $\rho _{g}$ in numerical studies aiming to
reproduce observations. When the depth of the atmospheric perturbation is $H$%
, the Rossby radius is $\rho _{g}=\left( gH\right) ^{1/2}/f_{0}$ (where $g$
is the gravitational acceleration). Actually $\rho _{g}$ is influenced by
several other parameters than $H$, in particular the vorticity and it can
have a range from tens to thousand kilometers. When combined with the range
of vortex extensions $R_{\max }^{phys}$, it results that $R_{\max }$ has a
range between a fraction of unity to few units (in plasma $R_{\max }$ can be
hundreads to thousands). For mature stationary cyclones the radial extension
of the cyclon vortex and the Rossby radius are of similar magnitude and this
means that we have to study the range $R_{\max }\sim 1$.

\bigskip

The domain of integration is a square with a side of length $2L$ 
\begin{equation}
\left( x,y\right) \in \left[ -L,L\right] \times \left[ -L,L\right]
\label{domain}
\end{equation}
on which we place a rectangular mesh $n\times n$ usually with $n=101$.

We impose Dirichlet boundary conditions \emph{i.e.} the streamfunction is a
constant, $\psi _{b}$ on the limits of the square of integration. Here $\psi
_{b}$ is the smaller root of the equation $\cosh \psi -p=0$ and since in all
our runs we have $p=1$, the condition is $\psi _{b}=0$. This means zero
vorticity at the boundary but it has defavorable effect on the velocity
field, which does not allways vanish at the border.

\bigskip

In general the initial profiles have been of various types: (a) symmetric
profiles (\emph{e.g.} Gaussian functions, or various annular shapes) with
maximum centered on $\left( 0,0\right) $, or the Petviashvili-Pokhotelov
vortex, Eq.(\ref{petv}) below; (b) functions expressed as product of
trigonometric functions; or (c) collections of localised vortex-like
perturbations placed randomly. In this work we report results obtained for
initializations with vorticity of the same sign over all the space region.
It resulted that the monopolar vortex which may be relevant for the
atmosphere physics can actually be obtained as a solution of the convergent
iteration from a large class of initial conditions, of various shapes
belonging to the three classes mentioned above. However, in terms of a set
of parameters describing any of the classes of initialization, finding the
convergence (solution) proves to be complicated and finding a solution is
rather exceptional. It can be said that the regions favorable to convergence
in the space of initial function are intricate and have sharp limits.

In order to simulate the formation of a vortex the main series of runs
reported here have adopted as initial function (from which the iteration of
GIANT starts) a monopolar, circularly symmetric form characterized by few
parameters, the Petviashvili-Pokhotelov vortex 
\begin{equation}
\psi \left( x,y\right) =\psi _{0}\left[ \mathrm{\sec h}\left( kr\right) %
\right] ^{4/3}+\psi _{b}.  \label{petv}
\end{equation}
Here $kr=\sqrt{\left( x^{2}+y^{2}\right) /\delta ^{2}}$, $\delta $ may be
seen as a peaking factor of the initial shape, $\psi _{0}$ is the amplitude
and $\psi _{b}$ is the boundary value. The reason to choose the form (\ref%
{petv}) is connected with the set of equation (continuity and conservation
of momentum) from which the CHM equation is originally derived. It has been
shown by a multiple space-time scale analysis that the late stage evolution
of the full set of equation is dominated by mesoscopic scales (of the order $%
\sqrt{L_{n}\rho _{s}}$ where $L_{n}$ is a typical length of the gradient of
the equilibrium density) where a different nonlinear mechanism is present.
Instead of the nonlinearity term in (\ref{CHM}) there is a scalar or
Korteweg-DeVries nonlinearity of the type $\psi \frac{\partial \psi }{%
\partial x}$ in one dimension, and the equation actually is replaced by the
Flierl-Petviashvili equation (see for example Spineanu et al. 2004). Or,
this equation has a solution expressed as Eq.(\ref{petv}). As initial
function, Eq.(\ref{petv}) has some advantages: it has physical relevance for
the system which is behind the stationary states emerging from Eq.(\ref{CHM}%
); it is a vortical structure, as those which may be expected to form
spontaneously in real situations; and it has few parameters which we take as
coordinates in a space of initial functions that must be sampled to find
solutions.

The parameters of the equation and of the initial function, which completely
defines a numerical experiment, are: (a) half the length of the side of the
square area taken as region of integration, $L=L^{phys}/\rho _{g}$; (b) the
constant of the equation which in these runs is taken $p=1$. This implies
that the boundary condition for the streamfunction, which also means zero
vorticity, is $\psi _{b}^{\left( 1,2\right) }=\ln \left( p\pm \sqrt{p^{2}-1}%
\right) =0$; (c) the peaking on the center, described by $\delta $; (d) the
amplitude of the initial function $\psi _{0}$.

\bigskip

The choice of an initial amplitude $\psi _{0}$ favorable for reaching a
solution is made easier if we use the following formula connecting the
radius of the maximum azimuthal velocity $a$ with the amplitude of the
streamfunction at this maximum $\psi _{0}$: 
\begin{equation}
a\sim \sqrt{\psi _{0}}\exp \left( -\psi _{0}+1\right)  \label{apsi0}
\end{equation}
This formula can be derived by simple manipulations of the equation (\ref{eq}%
), together with approximations that are made possible by some numerical
experience about the orders of magnitude of the normalized quantities. It is
a rather poor but useful approximation and only works for $L\lesssim 3$, a
range which is interesting for the atmospheric vortex.

From the other types of initial profiles we briefly discuss the second one.
As suggested by the numerical study of the \emph{sinh}-Poisson equation the
initial function has been taken in several runs as a product of
trigonometric functions in both directions, $x$ and $y$. To have a good
initialization, we choose a point $\left( x,y\right) $ where the initial
function is maximum. The equation imposes a condition on only the amplitude $%
\psi _{0}$ of the trigonometric functions of period $k$, 
\begin{equation}
\Delta \psi =\psi _{0}\left[ 2\left( k\pi \right) ^{2}\right] \simeq \frac{1%
}{2}\sinh \psi _{0}\left( \cosh \psi _{0}-1\right)  \label{guess}
\end{equation}
The equation is solved graphically and one of the roots is selected as the
amplitude of the initial function. From these initializations we obtain
sometimes quasi-solutions consisting of multiple vortices.

\subsection{Numerical solution for circularly symmetric vortices}

When the integration with GIANT identifies a monopolar, circularly symmetric
solution (usually with the precision $10^{-8}$) the shape is still
influenced by the square geometry of the domain of integration (\ref{domain}%
) and by the boundary condition $\psi _{b}=0$. The velocity is usually not
zero on the boundaries and this means that we have a too approximative value
for the total energy (this is not the case with the vorticity field and with
the energy and vorticity for the strongly localized vortices). In these
cases we can make a one dimensional (radial) integration 
\begin{equation*}
\frac{d^{2}\psi }{dr^{2}}+\frac{1}{r}\frac{d\psi }{dr}+\frac{1}{2p^{2}}\sinh
\psi \left( \cosh \psi -p\right) =0
\end{equation*}%
with $\psi =0$ at $r=R_{\max }$. Since the square and radial problem have
different boundary conditions the mapping between the solutions obtained in
square and in radial integrations requires certain care. We have found that
the general rule 
\begin{equation}
L=R_{\max }/\sqrt{2}  \label{elrmax}
\end{equation}%
establishes a good correspondence between the square solutions and the
radial solutions. This simply means that the solution on a square with half
side $L^{sq}\equiv L$ are very close to the radial solution with extension $%
L^{rad}\equiv R_{\max }$ equal to half the diagonal of the square. This is
shown by the Table 1 where we compare the important quantities: $%
r_{v_{\theta }^{\max }}$ (the radius where the maximum tangential velocity
is attained) and $v_{\theta }^{\max }$ (the maximum tangential velocity),
obtained in square (\textquotedblleft sq\textquotedblright ) and
respectively radial (\textquotedblleft rad\textquotedblright ) integrations.
These quantities are also compared in figures 1a and 1b.

\begin{center}
\begin{tabular}{|l|l|l|l|l|l|l|l|}
\hline\hline
$L^{sq}$ & $L^{rad}$ & $error^{rad}$ & $\left( r_{v_{\theta }^{\max
}}\right) ^{sq}$ & $\left( r_{v_{\theta }^{\max }}\right) ^{rad}$ & $\left(
v_{\theta }^{\max }\right) ^{sq}$ & $\left( v_{\theta }^{\max }\right) ^{rad}
$ & $\left( v_{\theta }^{perif}\right) ^{rad}$ \\ \hline\hline
$0.5$ & $0.707$ & $4.88$ & $0.05$ & $0.045$ & $21.44$ & $21.77$ & $2.26$ \\ 
\hline
$0.65$ & $0.92$ & $3.1$ & $0.07$ & $0.078$ & $13.18$ & $13.44$ & $1.67$ \\ 
\hline
$0.75$ & $1.06$ & $2.47$ & $0.095$ & $0.09$ & $10.8$ & $10.077$ & $1.42$ \\ 
\hline
$1.$ & $1.41$ & $1.48$ & $0.152$ & $0.16$ & $5.93$ & $5.91$ & $1.005$ \\ 
\hline
$1.25$ & $1.77$ & $0.95$ & $0.223$ & $0.225$ & $3.94$ & $3.85$ & $0.756$ \\ 
\hline
$1.5$ & $2.12$ & $0.71$ & $0.288$ & $0.33$ & $2.83$ & $2.83$ & $0.594$ \\ 
\hline
\end{tabular}
\end{center}

For radial integrations we use \emph{BVPLSQ} (Boundary Value Problem Least
Square Solvers for highly nonlinear equations), written by P. Deulfhard and
G. Bader. The code is a part of the \emph{CodeLib} library of Fortran codes
supported by Konrad Zuse Zentrum fuer Informationtechnik Berlin (ZIB) (see
the Prolog of BVPLSQ at the \emph{CodeLIB} web site). Although is more
efficient than \emph{MULCON} (also from \emph{CodeLIB}) or the NAG
subroutine \emph{D02GAF}, we have noted that the radial integration has a
weaker ability of identification of the solution compared with the
two-dimensional (square) one, GIANT. The solution is obtained at a lower
accuracy which we quantify by defining a functional $error^{rad}$%
\begin{equation}
error^{rad}=\int d^{2}r\left[ \left| \omega \right| -\left| NL\right| \right]
^{2}  \label{error}
\end{equation}
where $\omega $ is the vorticity and $NL$ is the nonlinear term in Eq.(\ref%
{eq}). This is shown in a column of the Table.

\begin{figure}[tbp]
\centering
\includegraphics[width=5cm]{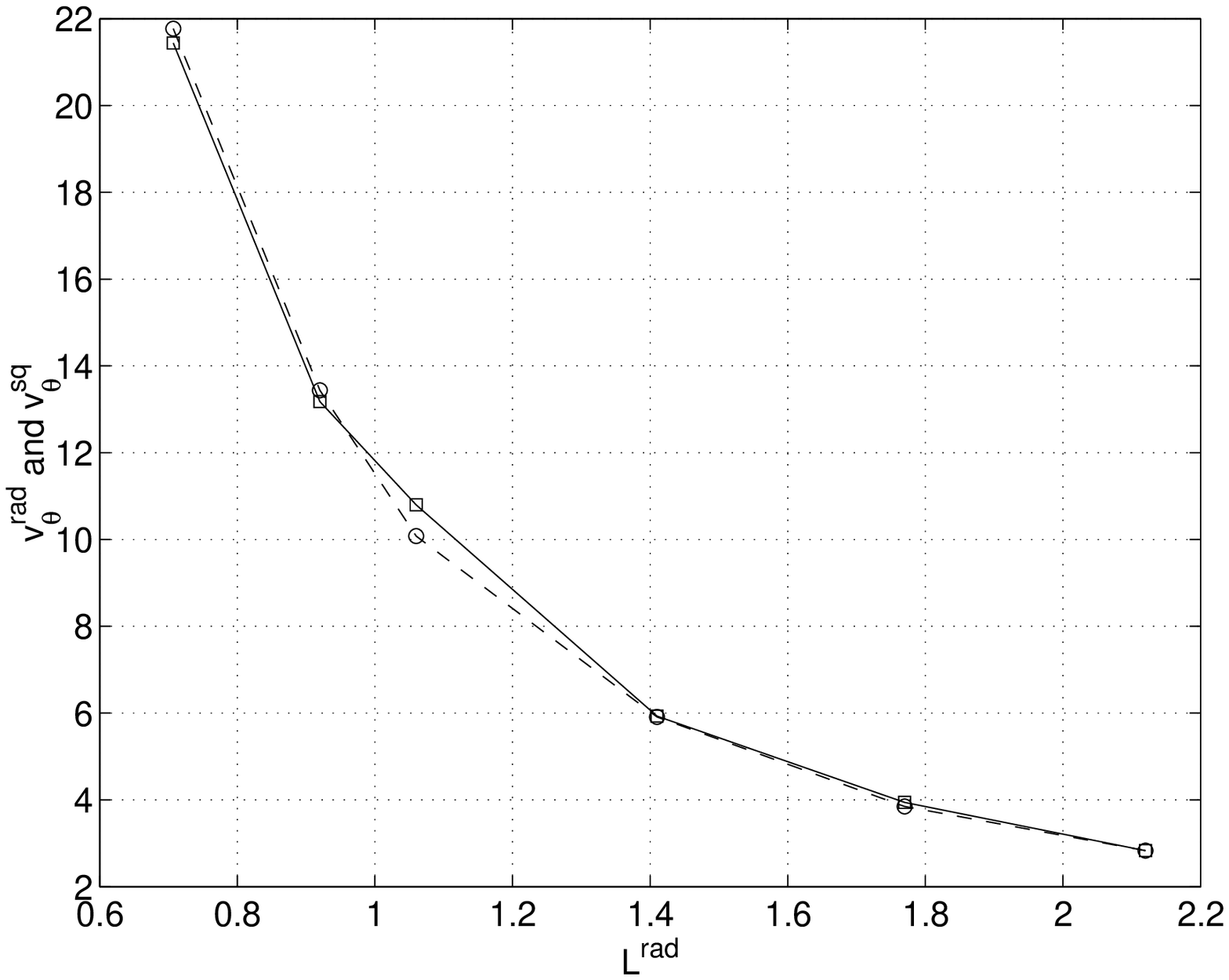}\hspace{0.1cm}%
\includegraphics[width=5cm]{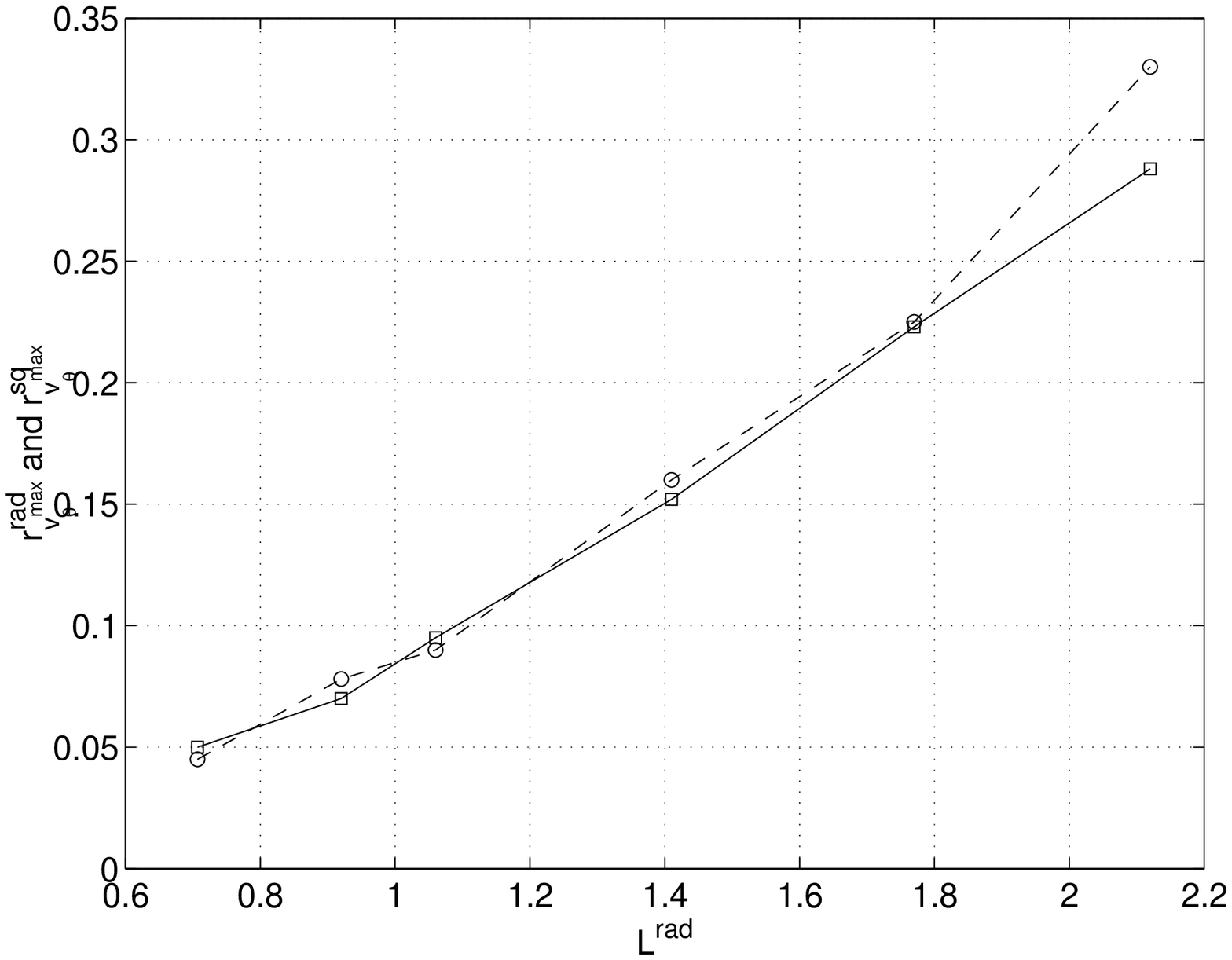}
\caption{Comparison between results on square (squares and continuous line)
and radial (circles and dashed line) integrations for: (a) $v_{\protect%
\theta }^{\max }$ and (b) $r_{v_{\protect\theta }^{\max }}$ }
\label{fig_1}
\end{figure}
A massive series of (automatic) radial integrations has been performed,
since they are faster than GIANT. Both solutions and quasi-solutions are
obtained and it is confirmed that the very precise results of the square
integrations (GIANT) are points on a line of minimum error (\ref{error}).
They will be reported elsewhere.

\section{Results}

\subsection{Summary of the numerical results}

The strong nonlinear character of Eq.(\ref{eq}) combined with the internal
procedures of GIANT (with no physical significance) imprints a particular
structure to a space of functions that is explored for exact solutions. The
space of functions representing initial conditions are divided into disjoint
parts, such that from one subset one cannot access the final configuration
of another subset. This means that in order to obtain a particular type of
stationary (asymptotic) solution one has to initialize in a particular
subset, homotopically connected to the final state. Most of the initial
conditions does not lead to convergence and possibly they correspond to
turbulent physical states. Exact solutions are obtained in the form of
trivial ($\psi \equiv 0$) state and monopolar or multipolar vortices.

Besides exact solutions there are sets of functions that are almost
solutions, \emph{i.e.} velocity field configurations that verify the
equation with only low accuracy and are normally rejected by the integration
procedure at smaller tolerance. They are interesting because they appear
systematically and approximately exhibit the same characteristics for a
fixed $L$. The iteration of GIANT gets almost stuck around such a solution,
which may suggest that they are metastable states of the physical fluid and
eventually evolve slowly toward an exact solution, a smooth vortex. We call
them quasi-solutions and we find useful to include them in our discussion.
The main reason for accepting them as interesting and possibly physically
relevant structures resides in the particularity of our approach: the
fundamental object is the action functional and the configurations described
by Eq.(\ref{eq}) extremize this action. If, by indifferently what method
(here numerical), it has been possible to identify a configuration that
seems to be very close to the extremum of the action (possibly a local
minimum in the function space) then this configuration may play a role in
the system's evolution. Although a more detailed description of the
structure of the function space around the extrema of the action is still
required, the highly concentrated vortices identified numerically seem to be
close of extremizing the action functional.

Restricting to the case of monopolar vortices we summarize the results by
saying that for every $L$ we obtain two types of final vortices:

\begin{enumerate}
\item smooth, finite amplitude, vortices verifying the equation for any
change in the accuracy of the integration procedure. There is a unique
smooth vortex configuration for each $L$.

\item quasi-solutions, consisting of strongly localised vortices, with
profiles for velocity and vorticity that are much narrower compared with the
smooth vortices. These quasi-solutions are persistently obtained and they
exhibit a clusterization around a typical shape for a particular $L$.
\end{enumerate}

\subsection{Solutions with the morphology of \emph{tropical cyclones}}

We are focusing here on the type of solutions that exhibit strong
similarities with the morphology of a horizontal cross section of a tropical
cyclone. These are circularly symmetric solutions with a strong maximum of
the tangential velocity, strong concentration of vorticity.

\begin{figure}[tbph]
\centerline{\includegraphics[height=5cm]{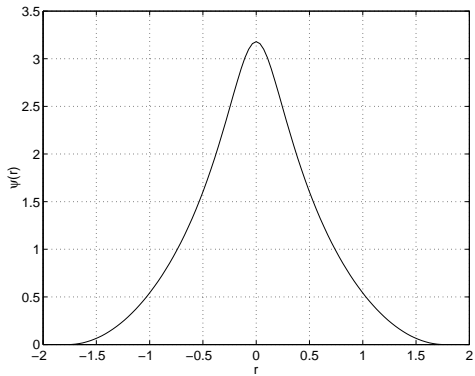}}
\caption{Solution $\protect\psi (x,y)$ along the diagonal of the square, for
the smooth vortex at $L=1.25$.}
\label{fig_2}
\end{figure}

We consider for ilustration the smooth, circularly symmetric vortex obtained
for $L=1.25$. Fig.\ref{fig_2} presents a section of the streamfunction $\psi
\left( x,y\right) $ along the diagonal of the domain $\left( x,y\right) $. A
section of the vorticity $\omega \left( x,y\right) $ is presented in Fig.\ref%
{fig_3}.

\begin{figure}[tbph]
\centerline{\includegraphics[height=5cm]{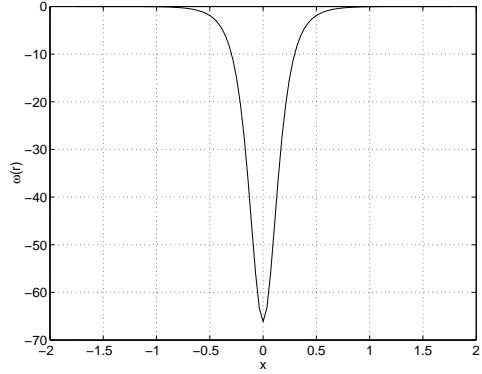}}
\caption{Vorticity calculated from $\protect\psi (x,y)$ obtained by
integration.}
\label{fig_3}
\end{figure}
\bigskip

\begin{figure}[tbph]
\centerline{\includegraphics[height=5cm]{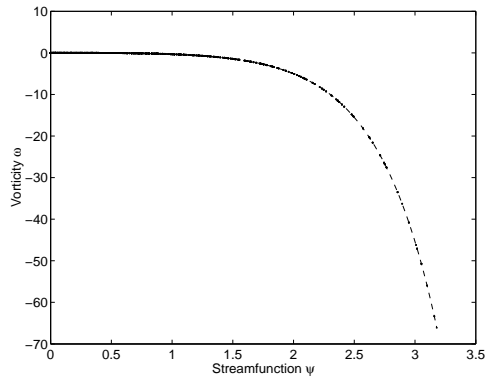}}
\caption{Scatterplot $(\protect\psi ,\protect\omega )$, for the smooth
vortex at $L=1.25$.}
\label{fig_4}
\end{figure}

In order to quantify the accuracy of integration we collect in all the
domain $\left( x,y\right) $ the pairs $\left( \psi ,\omega \right) $ and
plot them together with the line representing the nonlinear term in Eq.(\ref%
{eq}), Fig.\ref{fig_4}. The scatterplot of $\left( \psi ,\omega \right) $ is
almost superposed on this line. The scatterplot of the pairs [$\omega $, $-%
\frac{1}{2}\sinh \psi \left( \cosh \psi -1\right) $] (not shown) indicates a
close clustering around the diagonal. Other tests are possible and they show
that the integration is very good on most of the region and good within the
imposed accuracy in the regions where the second derivative is very high.

\begin{figure}[tbph]
\centerline{\includegraphics[height=10cm]{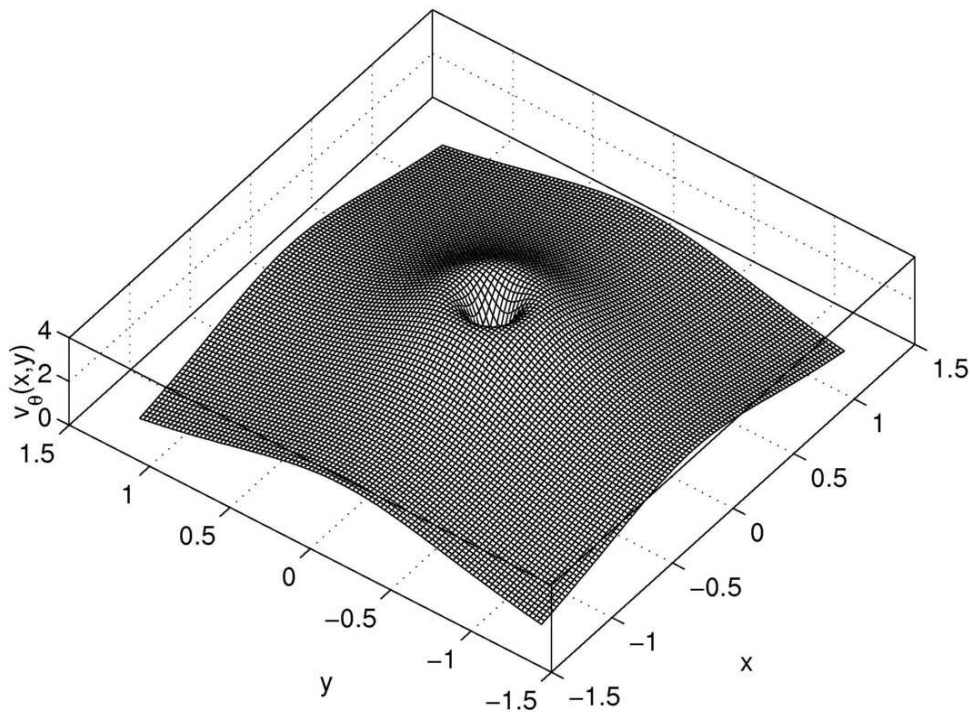}}
\caption{Azimuthal velocity $v_{\protect\theta }(x,y)$ for the smooth vortex
at $L=1.25$.}
\label{fig_5}
\end{figure}

The tangential component of the velocity is shown in Fig.\ref{fig_5}. The
steep descent to the center is clearly visible and its radial extension can
be compared with the extension of the whole domain.

\begin{figure}[tbph]
\centerline{\includegraphics[height=5cm]{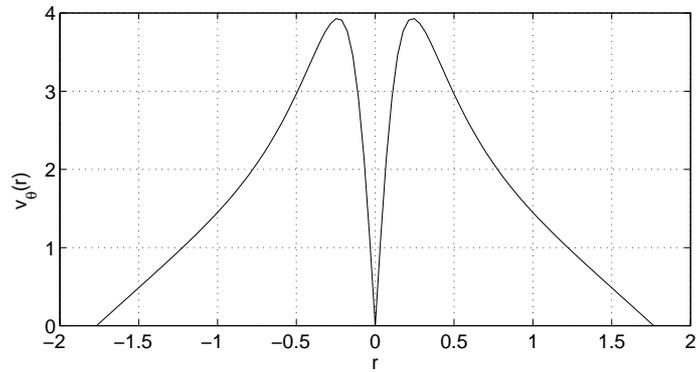}}
\caption{Azimuthal velocity $v_{\protect\theta }(x,y)$ along the diagonal of
the square of integration, for the smooth vortex at $L=1.25$.}
\label{fig_6}
\end{figure}

We have ploted in Fig.\ref{fig_6} the section along the diagonal of the
amplitude of the azimuthal component of the velocity.



\bigskip 
\begin{figure}[tbph]
\centerline{\includegraphics[height=5cm]{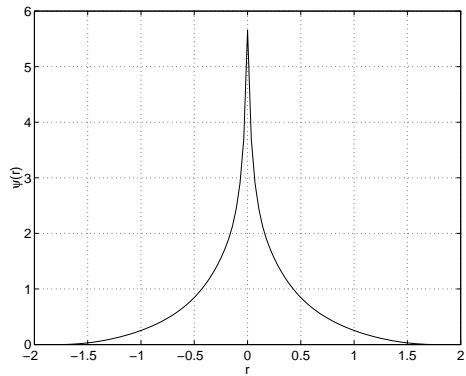}}
\caption{Streamfunction $\protect\psi (x,y)$ along the diagonal of the
square of integration, for the quasi-solution at $L=1.25$.}
\label{fig_7}
\end{figure}
\begin{figure}[tbph]
\centerline{\includegraphics[height=5cm]{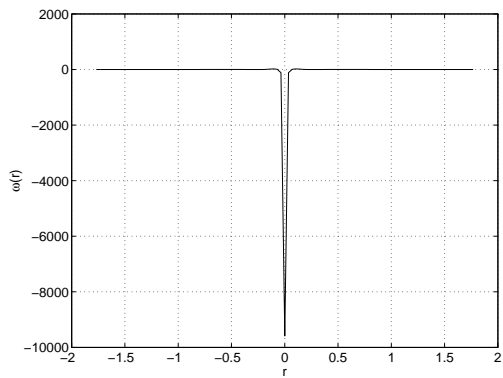}}
\caption{Vorticity $\protect\omega (x,y)$ along the diagonal of the square
of integration, for the quasi-solution at $L=1.25$.}
\label{fig_8}
\end{figure}
\begin{figure}[tbph]
\centerline{\includegraphics[height=5cm]{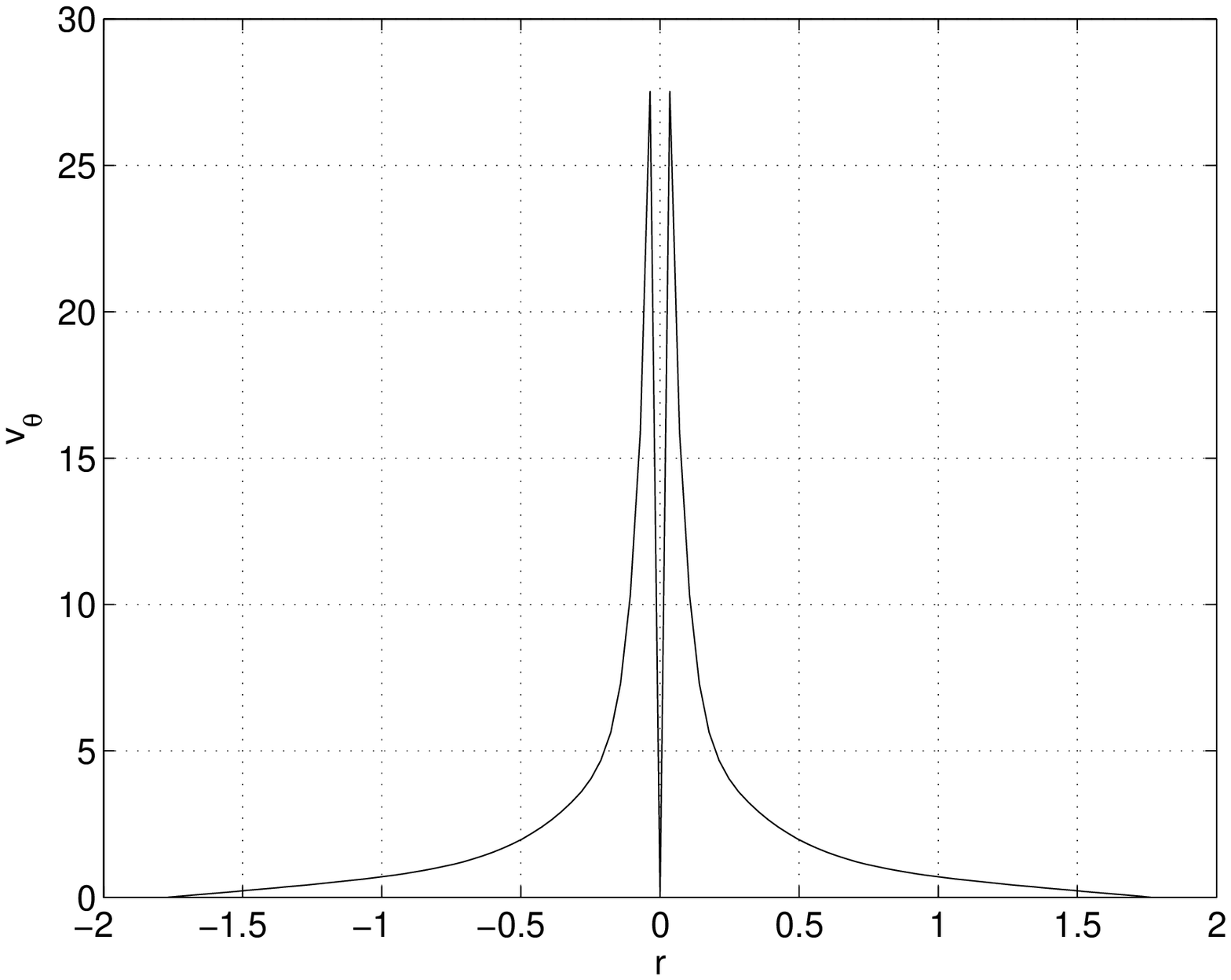}}
\caption{Azimuthal velocity $v_{\protect\theta }(x,y)$ along the diagonal of
the square of integration, for the quasi-solution at $L=1.25$.}
\label{fig_9}
\end{figure}

In general the radial component of the velocity is much smaller than the
azimuthal component. For this example ($L=1.25$) is in a ratio $\left|
v_{r}\right| /\left| v_{\theta }\right| \sim 0.4/4=1/10$ and integrated over
a circle shows no net inflow to the axis.

\bigskip

The large amount of results for the range of $L$ : $0<L\leq 10$, allows to
formulate two remarks. First we note that for larger $L$ the profile of the
azimuthal velocity shows smaller amplitude and larger radius of the circle
of maximum velocity. Second, this variation with $L$ is much faster for low $%
L$ (less or comparable to $1$). The differences in the quantitative
characteristics of the vortices for even small variation of $L$ in this
range are substantial. Below we provide a scaling which shows exponential
behavior, Eq.(\ref{vthmaxL}).

\subsection{Quasi-solutions : strongly localised vortices}

For comparison we present the profiles of the solution $\psi \left(
x,y\right) $ Fig.\ref{fig_7}, vorticity $\omega \left( x,y\right) $ Fig.\ref%
{fig_8} and azimuthal velocity $v_{\theta }\left( x,y\right) $ Fig.\ref%
{fig_9} for the quasi-solution that correspnds to the same $L=1.25$. The
sections are along the diagonal, denoted $r$. The results for every $L$'s
seem to indicate a clusterization of the final total energy and final total
vorticity. One should note that the total vorticity (obtained by integrating
over the square) is however smaller than that for the corresponding smooth
vortex shown in the preceding figures. The total energy $\frac{1}{2}\rho
_{0}\int d^{2}r\left[ \left( \mathbf{\nabla }\psi \right) ^{2}+\frac{1}{\rho
_{s}^{2}}\psi ^{2}\right] $ for a fluid density $\rho _{0}$, is larger for
the concentrated vortices compared to the smooth ones.

\subsection{Quasi-solutions : multiple vortices}

There are episodic structures of multiple vortices that are detected as
solutions under a certain precision and which however evlove to symmetric
monopolar vortex when the system is allowed to run further, under a higher
precision.

It is worth to mention that in a numerical experiment we have identified a
state where two vortices have been formed, placed in symmetrical positions
along the diagonal of the square domain $L=0.5$. The initial function is
trigonometric with periodicity $k = 2$ with a coefficient $\psi _{0}=3.8 $.
Examining this structure with higher precision, after a longer iteration
sequence the final solution was again the centered smooth vortex known for $%
L=0.5$. Therefore from the point of view of the numerical experience this
state of two vortices is irrelevant. However, the persistence of this state
inside the iterative search may indicate that it is close to a solution,
possible less structurally stable.

Four vortices have been obtained in a run starting from trigonometric
initial function. The initial function is trigonometric with periodicity $%
k=3 $. The results show the formation of four vortices, as shown by Fig.\ref%
{fig_10}. Each of them has a structure that is similar to the one presented
in Fig.\ref{fig_2}. It is interesting to note that again the vorticity is
almost zero everywhere on the domain, except the regions of the four
vortices, where it reaches very high values. The local tangential velocity
presents the same very fast decay to the center of the vortex and each
vortex is similar in structure with a typical cyclone.

\begin{figure}[tbph]
\centerline{\includegraphics[height=7cm]{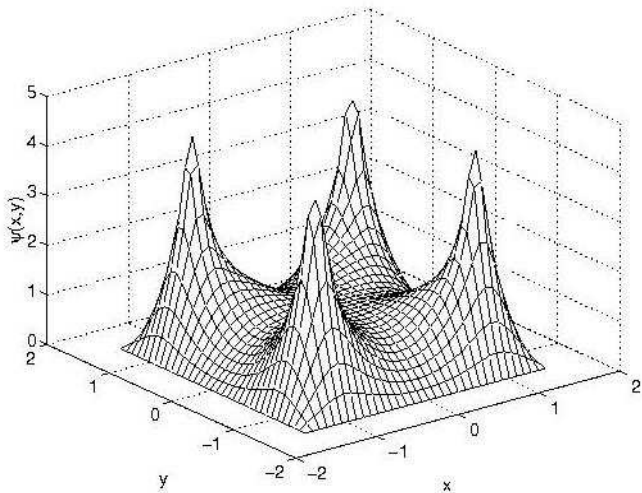}}
\caption{The scalar streamfunction $\protect\psi (x,y)$ for a four-vortices
solution.}
\label{fig_10}
\end{figure}
\begin{figure}[tbph]
\centerline{\includegraphics[height=7cm]{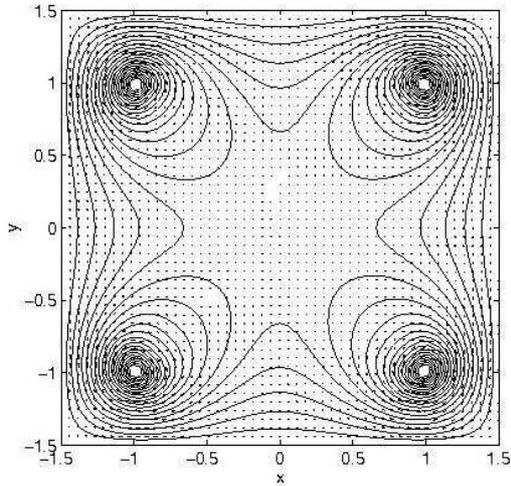}}
\caption{The contours of the scalar streamfunction $\protect\psi (x,y)$ and
the vector field $(v_{x}.v_{y})$ for a four-vortices solution.}
\label{fig_11}
\end{figure}

\section{Scaling laws inferred from numerical results}

As mentioned before the nontrivial results of monopolar vortices are
systematically of two types: a smooth vortex solution and a strongly
localised quasi-solution. These are always the same for a fixed $L$. Their
characteristics strongly depends on $L$, especially for the lower part of
the range, where $L$ is few units or less.

Since the monopolar final states are independent of the initial conditions
from which we start and of the particular numerical method of solution
(GIANT) the relationships between the characteristics of the final vortices
are objective and reflect properties of the equation itself. In addition the
results (smooth vortex and narrow quasi-solution) are unique for a
particular $L$, suggesting we can collect all numerical results in a form of
nomograms or analytic (eventually spline functions) fit. However, we will
look instead for analytic formulas which, even approximative, are simpler to
use. We examine (a) the scaling of the \emph{maximum tangential velocity}
with the radius of the vortex, $R_{max}$; (b) the scaling of the \emph{%
radius of the eye-wall} of the atmospheric vortex with the length $L$.
Finally we will examine the existence of a linear relation between\emph{\
the energy} and \emph{the vorticty}in the final states.

\subsection{The relationship between the maximal tangential velocity and the
extension of the atmospheric vortex}

We have inferred from numerical data an expression showing variation of $%
v_{\theta }^{\max }$ with the radial extension of the vortex $R_{\max }$ 
\begin{equation}
v_{\theta }^{\max }\left( L\right) \simeq \frac{e^{2}}{2}\left[ \alpha \exp
\left( \frac{\sqrt{2}}{R_{\max }}\right) -1\right]  \label{vthmaxL}
\end{equation}
for the interval $0<L=R_{\max }/\sqrt{2}<6$. This is shown in Fig.\ref%
{fig_12}.

\begin{figure}[tbph]
\centerline{\includegraphics[height=7cm]{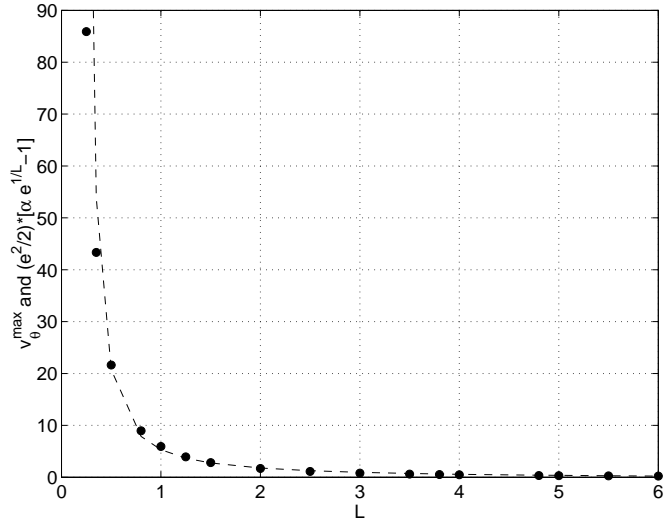}}
\caption{The maximum tangential velocity $v_{\protect\theta }^{max}$ as
function of $L=R_{\max}/\protect\sqrt{2}$. The dashed line represents the
fit according to Eq.(\protect\ref{vthmaxL})}
\label{fig_12}
\end{figure}

For $\alpha =0.97$ this scaling works better for $L$ around $1$, and is a
poor approximation over the range less than $0.4$. At $L\gtrsim 1.5$ this
formula overestimates the maximum velocity and will not be used when we
dispose, for the particular $L$, of the full numerical set as obtained by
GIANT. However many observational data fall in the range $L\sim 1$ where Eq.(%
\ref{vthmaxL}) is a good fit and there it may be useful for a rapid
estimation.

\subsection{The relationship between the radius of the eye-wall and the
extension of the atmospheric vortex}

On the basis of many runs we have tried to infer a possible relationship
between the radius of the \emph{eyewall}, (the radius of the circle where
the tangential velocity is maximum) and the parameter $L$.

When the data collected for a larger range of $L$, $0<L\leq 6$, is taken
into account, it appears that there is an approximative linear dependence of 
$r_{v_{\theta }^{max}}\left( L\right) $ on $L=R_{\max }/\sqrt{2}$. 
\begin{equation}
r_{v_{\theta }^{max}}\left( L\right) =0.11\left( -\frac{1}{2}+L\right) 
\label{eyefit1}
\end{equation}%
\begin{figure}[tbph]
\centerline{\includegraphics[height=7cm]{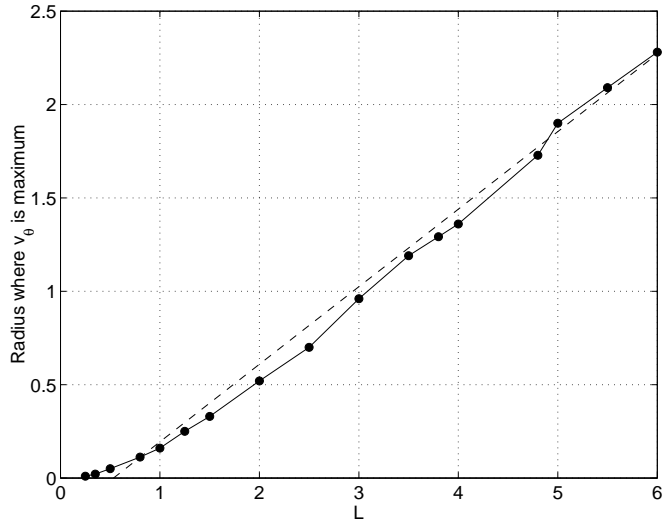}}
\caption{The radial position $r_{v_{\protect\theta }^{\max }}$ where the
tangential velocity $v_{\protect\theta }^{max}$ attains its maximum, as
function of $L$. The dashed line represents the linear fit according to Eq.(%
\protect\ref{eyefit1}) }
\label{fig_13}
\end{figure}

As can be seen from Fig.\ref{fig_13} this linear fit may present interest
for a very wide range of $L$'s but it does not work well for low $L$, $%
0<L\lesssim 2.5$. Or this is the range that is relevant for the atmospheric
vortex. It is necessary to look for a different fit for that range.

\begin{figure}[tbph]
\centerline{\includegraphics[height=7cm]{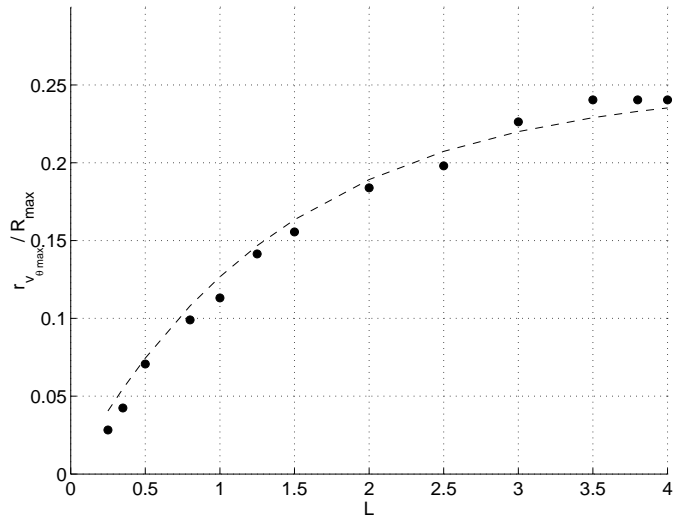}}
\caption{Ratio of the radius $r_{v_{\protect\theta }^{\max }}$ where maximum
of $v_{\protect\theta }$ is attained to the radius of the maximum extension
of the vortex, $R_{max}$ represented as function of the length $L=R_{\max }/%
\protect\sqrt{2}$. The dashed line is the fit according to Eq.(\protect\ref%
{eyefit2}).}
\label{fig_14}
\end{figure}
First we normalize the eye-wall radius to the radial extension of the
vortex, $R_{max}\sim \sqrt{2}L$. Then the same numerical information can be
organized to show the dependence of $r_{v_{\theta }^{max}}/R_{max}$ on the
length $R_{\max }$. The following simple function offers a satisfactory fit
for the low $L$ range 
\begin{equation}
\frac{r_{v_{\theta }^{max}}}{R_{max}}=\frac{1}{4}\left[ 1-\exp \left( -\frac{%
R_{max}}{2}\right) \right]   \label{eyefit2}
\end{equation}%
Although it slightly overestimates the ratio (see Fig.\ref{fig_14}), this
formula is practical by its simplicity and may be used for estimations based
on observational data, as will be described later.

\subsection{Energy and vorticity in the final states}

The smooth and respectively the strongly concentrated vortices are
persistently obtained, from a wide variety of initial conditions. This
suggests that, apart variations due to the accuracy of the iteration
process, these states really represent a stationary state and respectively a
quasi-stationary state of the fluid. Their characteristics are completely
determined once we have fixed the radial extension of the domain $R_{\max }$
or, for square integration, the length $L$, which is half the side of the
square on which the integration is performed, with $\psi =0$ on its
boundaries. 
\begin{figure}[tbph]
\centerline{\includegraphics[height=7cm]{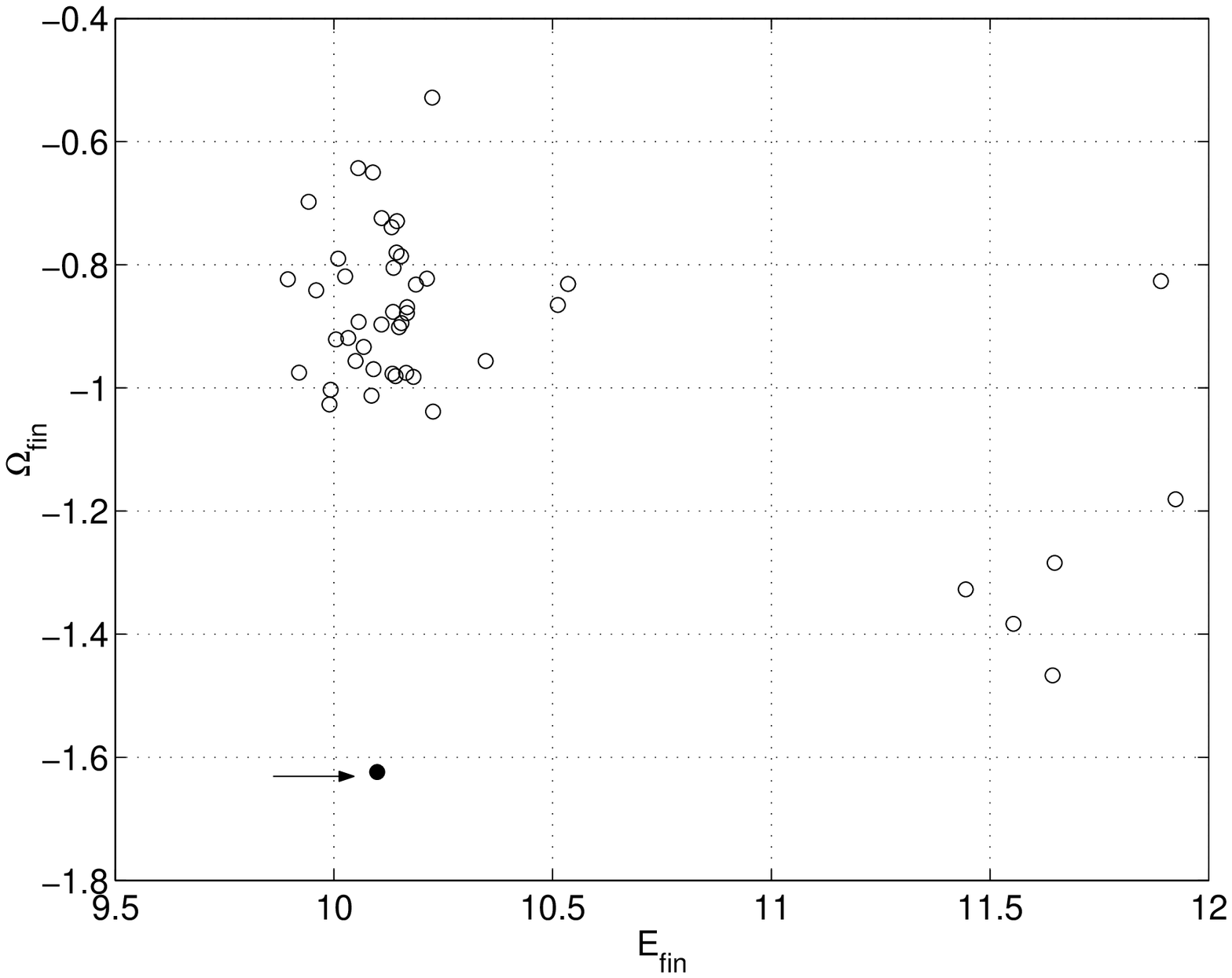}}
\caption{Energy and vorticity in the final states obtained at $L=1.25$. The
dot indicated by the arrow consists of $84$ smooth vortices, while the $46$
narrow vortices (open circles) show a dispersion in energy and vorticity.}
\label{fig_15}
\end{figure}
\begin{figure}[tbph]
\centerline{\includegraphics[height=7cm]{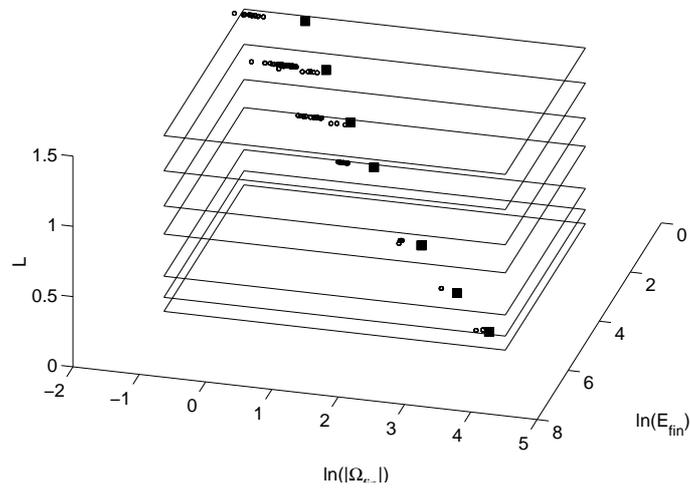}}
\caption{Energy and vorticity in the final states for several low $L$
values. The smooth (black squares) and concentrated vortices (small circles)
are plotted in logarithmic scale. Every plane corresponds to a particluar $L$
value: $0.25$, $0.35$, $0.5$, $0.8$, $1.0$, $1.25$ and $1.5$.}
\label{fig_16}
\end{figure}

In particular the final states for a fixed $L$ are characterised by two
quantities, the total final energy and the total final vorticity and there
are two pairs $\left( E_{fin},\Omega _{fin}\right) $: one for the smooth
vortex and one for the narrow vortex. They are calculated with 
\begin{eqnarray}
E^{phys} &=&\rho _{0}^{phys}c_{s}^{2}\left( R_{\max }^{phys}/\sqrt{2}\right)
^{2}  \label{e} \\
&&\times \frac{4}{n_{x}n_{y}}\frac{1}{2}\sum_{i=1}^{n_{x}}\sum_{j=1}^{n_{y}}%
\left[ \left\vert v\left( i,j\right) \right\vert ^{2}+\psi \left( i,j\right)
^{2}\right]   \notag
\end{eqnarray}%
where $\rho _{0}^{phys}$ is the density. We denote $E$ the energy $E^{phys}$
normalized to the physical coefficient in the first line above. The
vorticity is 
\begin{eqnarray}
\Omega ^{phys} &=&f_{0}  \label{o} \\
&&\times \frac{1}{n_{x}}\frac{1}{n_{y}}\sum_{i=1}^{n_{x}}\sum_{j=1}^{n_{y}}%
\omega \left( i,j\right)   \notag
\end{eqnarray}

When pairs $\left( E,\Omega \right) $are collected from all numerical
experiences for a particular $L$, the precision of numerical determination
of the solution is reflected in the dispersion of the points around an
average one which may be supposed to be the exact solution. For the smooth
vortices the dispersion is insignifiant, confirming that the numerical
identification of this solution is very good, for all initializations. For
the strongly localized vortex the dispersion is apparent and is connected
with the higher magnitude of the second derivative on a very small area.
However the results are clearly clustered around an average value which one
can use to study various scaling relationships. This difference is
exemplified in Fig.\ref{fig_15} showing $130$ results for $L=1.25$, both
smooth and narrow vortices, obtained from widely different initial
conditions. The black dot indicated by the arrow actually consists of $84$
distinct dots representing almost identical results for the smooth vortex at 
$L=1.25$. The open circles are $46$ quasi-solutions narrow vortices, which
have very close characteristics ($v_{\theta }\left( x,y\right) $, $%
r_{v_{\theta }^{\max }}$,etc.). They seem to represent approximations of a
unique quasi-solution. The dispersion in the final energy and vorticity are
a consequence of differences in the calculated second derivative of $\psi $
on a very small area, the reason for which they are actually rejected by
GIANT. In Fig.\ref{fig_16} data are plotted for several values of $L$.

\begin{figure}[tbp]
\centering
\includegraphics[width=6cm]{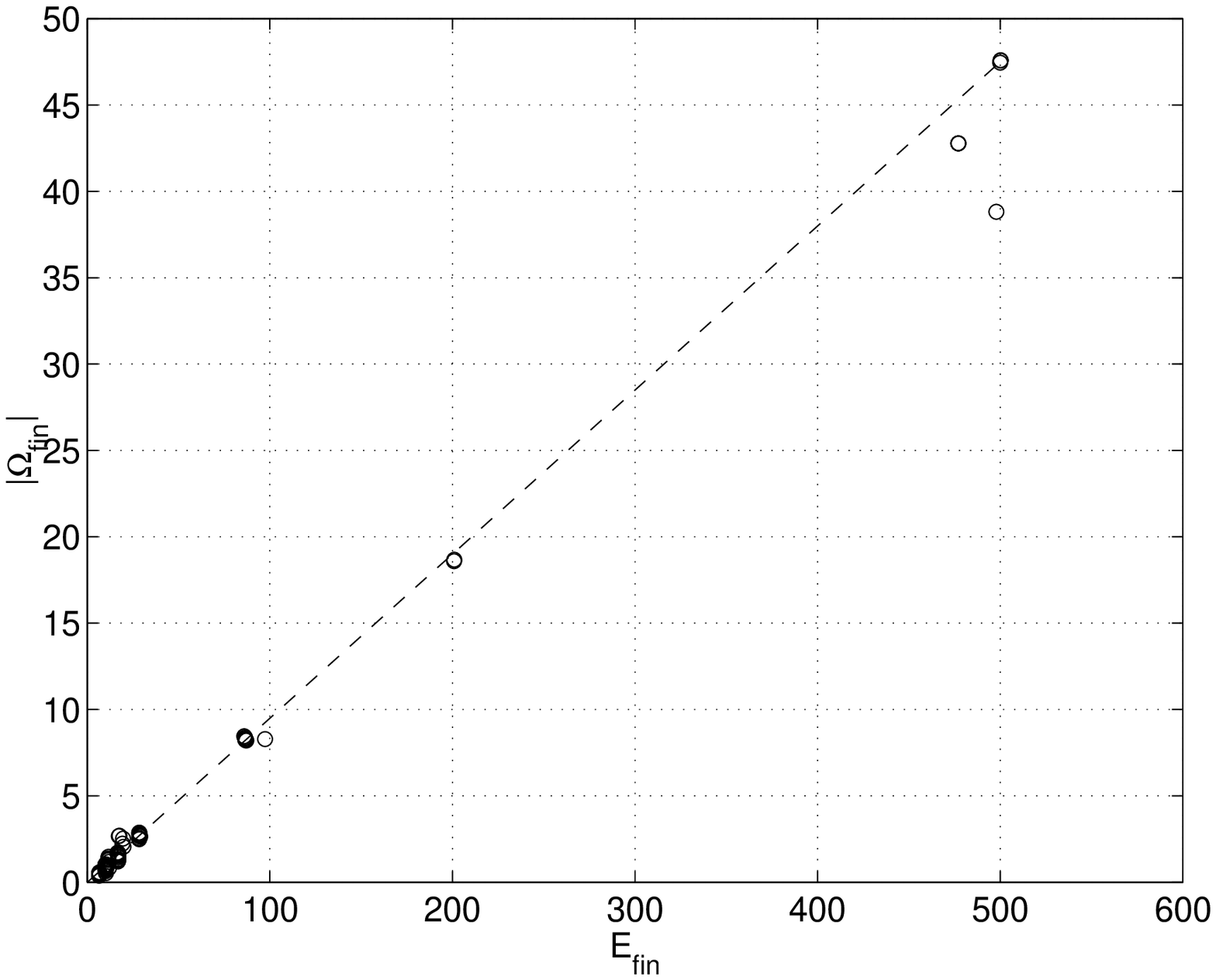}\hspace{0.5cm}%
\includegraphics[width=6cm]{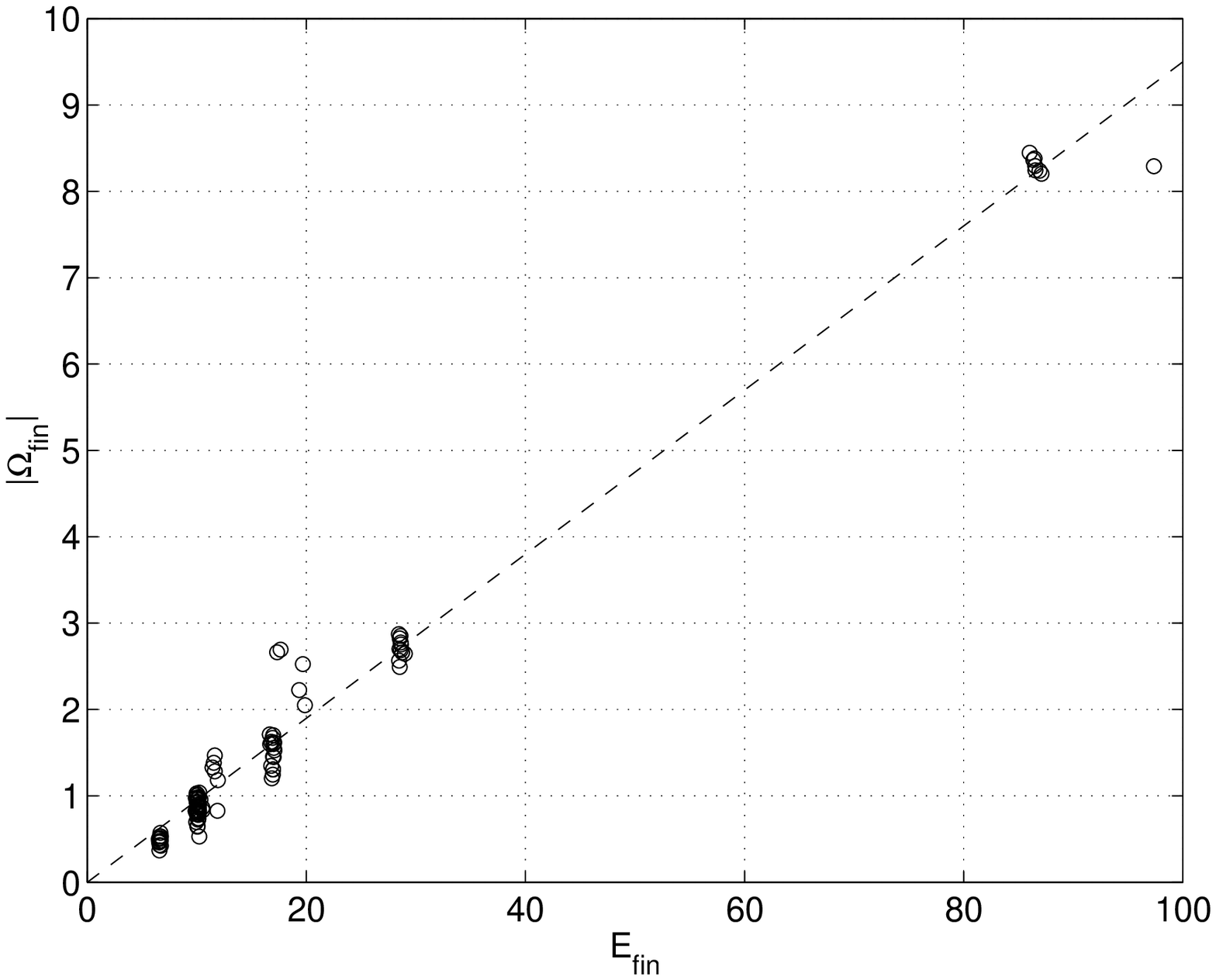}
\caption{(a) Energy and vorticity in the final states corresponding to
strongly localized vortices. The dashed line shows that there is a linear
relationship between $E_{fin}$ and $\Omega_{fin}$. The figure (b) expands
the region close to the origin.}
\label{fig_17}
\end{figure}

The first important relation is between the energy $E_{fin}$ and the
vorticity $\Omega _{fin}$ in the final states. The field theoretical model
from which the equation is derived points out the existence of a lower bound
for the energy functional for the point-like vortices: the energy is
expressed as a sum of squared terms plus a supplementary term that has a
topological content. The minimum of the energy is obtained by taking the
squared terms to zero (and this leads to the self-dual equations and further
to Eq.(\ref{eq})) and this makes the energy equal to the topological term.
Integrating over all plane this equality takes the form of a proportionality
of the total energy and the total vorticity in the fluid motion. Therefore
the numerical results collected for all $L$ should exhibit a linear
relationship between $E_{fin}$ and $\Omega _{fin}$. Now we have for each $L$
two pairs $\left( E_{fin},\Omega _{fin}\right) $ and we can try to verify
this linear relations for both. The total vorticity can be calculated easily
in both cases, since the vorticity has a very good spatial limitation around
the eye and decays rapidly to zero. The total energy Eq.(\ref{e}) includes
the integration of the squared velocity but the square domain of integration
actually does not have everywhere zero velocity at the boundary, especially
for higher $L$. Then a certain amount of energy cannot be included and $%
E_{fin}$ is not reliable. This is the case with almost all smooth vortices
(except possibly the very small $L\sim 0.25...0.5$, where the localization
of the smooth vortex is more pronounced). However, for the strongly
localized vortices this problem does not arise, for any $L$. Fig.\ref{fig_17}
shows that there is indeed a linear relation between the energy and the
vorticity. With lower accuracy (which we have mentioned before) the radial
integrations confirm the proportionality of $E_{fin}$ and $\Omega _{fin}$.

\subsection{The existence of a threshold in the initial energy and vorticity
for obtaining a solution}

\begin{figure}[tbph]
\centerline{\includegraphics[height=7cm]{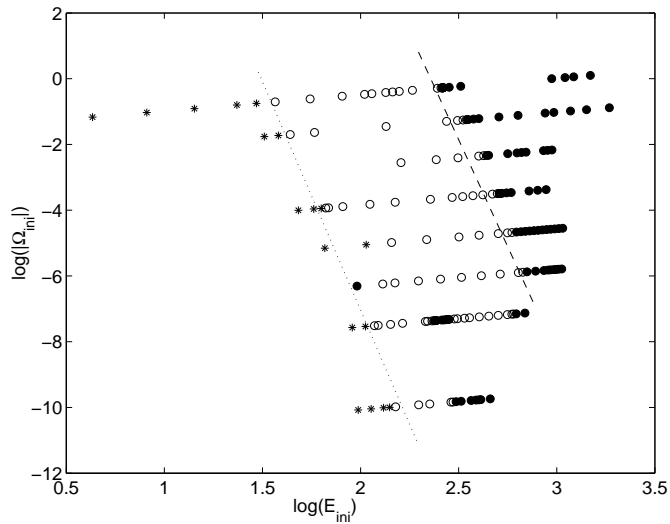}}
\caption{Set of points in the plane ($E_{ini}, \Omega_{ini}$) from which the
integrations for $L=1.25$ have been started, with the initial form Eq.(%
\protect\ref{petv}). The points are distinguished according to the final
states: (1) asterisks are initializations leading to trivial, zero, final
state; (2) open circles are initializations leading to smooth vortex
solutions; (3) black circles are initializations leading to quasi-solutions
with strongly localised vorticity. The dotted line is a tentative separation
of the trivial solutions (at left) from the smooth vortices (at right). The
dashed line is an orientative separation of smooth from the strongly
localised vortices. To the right of the righmost black circles there is no
solution.}
\label{fig_18}
\end{figure}

Numerical simulations of basic fluid equations have shown that there is a
boundary in the space of the initial configurations which separates two
kinds of behaviors: on one side there are states from which the fluid
evolves to random, turbulent states and on the other side there are initial
configurations giving in long run organised, highly ordered flow with a
vortical pattern. In the case of the Navier-Stokes equation this limit has
been called the \emph{ergodic boundary} by Deem and Zabusky (1971). In our
case there is no time evolution and the iterations of GIANT have no
particular physical meaning. We simply note however that a similar
separation occurs in the present case. For any fixed $L$ the iteration of
GIANT converges to a smooth vortex only if the amplitude of the initial
function $\psi _{0}$ is higher than a particular value, depending on the
peaking parameter $\delta $. This can be translated into a condition for the
initial energy and vorticity. Choosing $L=1.25$ we have represented in Fig.%
\ref{fig_18} the points $\left( E_{ini},\Omega _{ini}\right) $ corresponding
to all types of final results: zero (asterisk), smooth vortex (open
circles), narrow quasi-solution (black dots).

For purely orientative purpose a line is drawn, which approximately
separates initial states leading to trivial solution $\psi \equiv 0$ (at
left) from the initial states leading to smooth vortices (at right). This is
very steep, $\Omega _{ini}-\Omega _{ini}^{0}\sim \left(
E_{ini}-E_{ini}^{0}\right) ^{-13.3}$ with $\left( E_{ini}^{0},\Omega
_{ini}^{0}\right) $ one of the points on the limit. The other line is also
orientative, separating the initial conditions leading to smooth (at left)
and respectively strongly localised (at right) vortices. \FloatBarrier

\subsection{Radial profile of the azimuthal velocity: comparison with
Holland's model}

In an integration over square domain we obtain $\left( v_{x},v_{y}\right) $
with very good accuracy but the outer part of the field may be affected by
the square geometry. For monopolar vortices we repeat the integration in
radial geometry, which is extended over the length of the diagonal of the
square, as explained. This profile can be compared with semi-empiric
formulas like Holland's or DiMaria. For the Holland's model we use the
formula 
\begin{equation*}
v_{\theta }^{H}\left( r\right) =\left\{ \frac{b}{\rho _{0}}\left( \frac{%
r_{v_{\theta }^{\max }}}{r}\right)^b \left( P_{e}-P_{c}\right) \exp \left[
-\left( \frac{r_{v_{\theta }^{\max }}}{r}\right) ^{b}\right] +\frac{%
r^{2}f^{2}}{4}\right\} ^{1/2}-\frac{rf}{2}
\end{equation*}
The parameters are: $b=2$ (the shape parameter), $P_{c}=990\,\left(
hPa\right) $ (the pressure in the center of the vortex), $P_{e}=1015\,\left(
hPa\right) $ (the environmental pressure), $\rho _{0}=1.15\,\left(
kg/m^{3}\right) $, $f=5\times 10^{-5}\,\left( s^{-1}\right) $. The
parameters used in this example correspond to 
\begin{equation*}
\left( r_{v_{\theta }^{\max }}\right) ^{phys}\simeq 10\,\left( km\right)
\;,\;R_{\max }^{phys}=95\,\left( km\right)
\end{equation*}
and this can be taken as starting point of our calculations based on Eq.(\ref%
{eq}) and the scaling laws derived from it. We calculate the ratio of the
two distances $r_{v_{\theta }^{\max }}^{phys}/R_{\max }^{phys}=10/95=0.1053$%
. We insert this value in the scaling described by Eq.(15) and determine the
maximum radial extension of the vortex (normalised), $R_{\max }$%
\begin{equation*}
\frac{1}{4}\left[ 1-\exp \left( -\frac{R_{\max }}{2}\right) \right] =\frac{%
r_{v_{\theta }^{\max }}}{R_{\max }}=\frac{r_{v_{\theta }^{\max }}^{phys}}{%
R_{\max }^{phys}}=0.1053
\end{equation*}
From here we obtain 
\begin{equation}
R_{\max }=1.0931  \label{eq502}
\end{equation}
or, equivalently $L^{sq}=R_{\max }/\sqrt{2}=0.7729$. The Rossby radius, is 
\begin{equation*}
\rho _{g}=\frac{R_{\max }^{phys}}{R_{\max }}=\frac{95\;\left( km\right) }{%
1.0931}=86.9\;\left( km\right)
\end{equation*}
This is the first physical unit that we need in order to translate our
numerical results (normalised quantities) into physical quantities. The unit
of vorticity is $f_{0}=5\times 10^{-5}\;\left( s^{-1}\right) $ and the unit
of velocity 
\begin{equation*}
\overline{v}=\rho _{g}f_{0}=86.9\times 10^{3}\times 5\times 10^{-5}\simeq
4.34\;\left( m/s\right)
\end{equation*}

The numerical solution of Eq.(\ref{eq}), obtained with the code \emph{BVPLSQ}
in circular symmetry for $L^{rad}\equiv R_{\max }=1.0931$ , gives 
\begin{equation*}
v_{\theta }^{\max }\simeq 9
\end{equation*}%
This result is confirmed by the integration on the square domain $%
L^{sq}=R_{\max }/\sqrt{2}=0.7729$, with \emph{GIANT}, where we obtain $%
v_{\theta }^{\max }\simeq 9.5$. As explained before, the integration over
the square domain is more precise compared with the radial one (\emph{GIANT}
versus \emph{BVPLSQ}). However to compare with the profile of Holland's
model, we take the result of the \emph{radial} integration. We can now
calculate this velocity in \emph{physical} units 
\begin{equation*}
\left( v_{\theta }^{\max }\right) ^{phys}=4.34\,\left( m/s\right) \times
9\simeq 39.11\;\left( m/s\right) 
\end{equation*}%
\begin{figure}[tbph]
\centerline{\includegraphics[height=5cm]{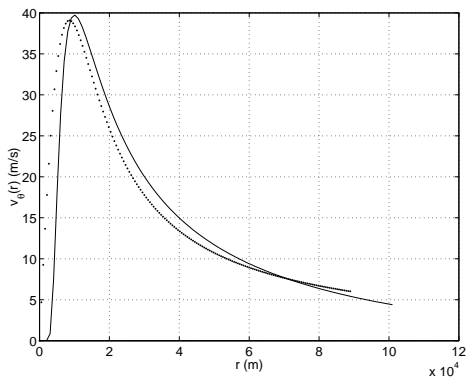}}
\caption{Azimuthal velocity from the Holland model (continuous line) and
from Eq.(\protect\ref{eq}) (circles).}
\label{fig_19}
\end{figure}

The Holland's model gives approximately $39.73\;\left( m/s\right) $. In
addition we plot in Fig.\ref{fig_19} the radial profiles of the azimuthal
velocity from the Holland's model (continuous line) and from our Eq.(1)
(dotted line), in physical units. The similarity of profiles is apparent. We
note however that these profiles are senzitive to choices that are difficult
to measure in observation: for the Holland's model the parameters, in
particular $b$, are affected by imprecisions. For our model, the main
physical input is the radius of the eye wall and the maximum extension of
the cyclone and these are equally affected by imprecisions.

\section{Practical application of the scaling laws}

In this section we discuss how to use the scaling relations we have derived
when we want to compare with a real observational data of a tropical
cyclone. Since for a fixed $L$ the integration provides a unique smooth
vortex, all data regarding this vortex are available in the form of
functions : $\psi \left( x,y\right) $, $\omega \left( x,y\right) $, $%
v_{\theta }\left( x,y\right) $, and derived quantities : $r_{v_{\theta
}^{\max }}$, $v_{\theta }^{\max }$, $E_{fin}$, $\Omega _{fin}$. All are 
\emph{normalized} and the first task is to identify the physical units that
will relate these quantities to the physical data. Basically there are two
units: $\rho _{g}$ and $f_{0}$ and the latter is assumed known.

A possible starting point is to estimate (from observations) the ratio
between the radius of the circle of maximum azimuthal velocity $r_{v_{\theta
}^{\max }}^{phys}$ and the radius of the minimum disc containing the
cyclone, $R_{\max }^{phys}$ (here as usual the upperscript \emph{phys} means
that the quantities are dimensional). Suppose this ratio can be estimated on
the basis of a satelite picture. Then we can use Eq.(\ref{eyefit2}) or Fig.%
\ref{fig_13} to identify the parameter $L$. Since a real-life observation
provides also the physical radius of the cyclone, $R_{\max }^{phys}$ we can
determine the Rossby radius from: $R_{\max }^{phys}=\sqrt{2}L^{phys}=\sqrt{2}%
\rho _{g}L$ where we use $L$ as determined. Then the unit of the physical
quantities are calculated : $\rho _{g}f_{0}$ for velocity and $\rho
_{g}^{2}f_{0}$ for the streamfunction. From the runs of the code for $L$ we
dispose of profiles for all (normalized) variables. Using the units we can
calculate some characteristics that can be further compared with the
observations: the maximum azimuthal velocity $v_{\max }^{phys}$, the profile
of the velocity, the vorticity, etc.

\subsection{Example 1}

We use pictures of the profile of the mean tangential wind for the hurricane
Andrew, according to Willoughby and Black (1996). In Fig.3b of this
reference it is represented the west to east wind profile before the eyewall
replacement (23 August 1665 UTC). We retrieve the approximate values: $%
v_{\theta \max }^{phys}\simeq 68\,\left( m/s\right) $, $r_{v_{\theta }^{\max
}}^{phys}\simeq 12\,\left( km\right) $ and we assume (with a certain
extension beyond the limits presented in the figure) $R_{\max }^{phys}\simeq
120\,\left( km\right) $. From this we calculate 
\begin{equation}
\frac{r_{v_{\theta }^{\max }}^{phys}}{R_{\max }^{phys}}\simeq \frac{12}{120}%
=0.1  \label{andrew_1}
\end{equation}%
With this value we turn to Eq.(\ref{eyefit2}) to calculate $R_{\max }$ and
then $L$. 
\begin{equation}
\frac{1}{4}\left[ 1-\exp \left( -\frac{R_{\max }}{2}\right) \right] =\frac{%
r_{v_{\theta }^{\max }}}{R_{\max }}=\frac{r_{v_{\theta }^{\max }}^{phys}}{%
R_{\max }^{phys}}=0.1  \label{andrew_2}
\end{equation}%
with the result 
\begin{equation}
R_{\max }\simeq 1.0217\;,\;L\simeq 0.72  \label{andrew_3}
\end{equation}%
Taking $L=0.72$, we have at this moment at our disposal all the set of
results that are obtained numerically for the smooth vortex at this $L$.
Coming back to the physical data we now use the spatial extension of the
vortex, $R_{\max }^{phys}\simeq 120\,\left( km\right) $ to calculate the
Rossby radius $\rho _{g}$, \emph{i.e.} the space normalization 
\begin{equation*}
\rho _{g}=\frac{R_{\max }^{phys}}{\sqrt{2}L}=\frac{120}{\sqrt{2}\times 0.72}%
\simeq 117.85\,\left( km\right) 
\end{equation*}%
Now we can calculate the other physical units: $\overline{\omega }%
=f_{0}=5\times 10^{-5}\;\left( s^{-1}\right) $ (from Emanuel 1989, Table 1),
for velocity $\rho _{g}f_{0}=5.9\,\left( m/s\right) $, for streamfunction $%
\rho _{g}^{2}f_{0}=0.694\times 10^{6}\,\left( m^{2}/s\right) $. By
numerically solving Eq.(\ref{eq}) for $L=0.72$ we find $v_{\theta }^{\max
}=10.9$. Note that the Eq.(\ref{vthmaxL}) gives a close value: $10.67$. The
physical value results $v_{\theta }^{\max }=64.31\,\left( m/s\right) $. This
is comparable to $68\,\left( m/s\right) $ the value shown by Willoughby and
Black. The maximum vorticity is $\left\vert \omega _{\max }\right\vert
=491.2\times f_{0}=0.0246\,\left( s^{-1}\right) $.

\subsection{Example 2}

For this example we adopt the following input data: the ratio of the full
spatial extension of the hurricane to the radius of maximum tangential wind
is $R_{\max }^{phys}/r_{v_{\theta }^{\max }}^{phys}\sim 9$; and the physical
extension of the hurricane is $R_{\max }^{phys}\sim 300\,\left( km\right) $.
The data can be compared with the picture taken by NASA at 28 August 2005,
when the hurricane Katrina was above the Mexic gulf, but the identification
of physical data is certainly approximative. Eq.(\ref{eyefit2}) is used to
find 
\begin{equation*}
L\sim \frac{1}{\sqrt{2}}\left( -2\right) \ln \left( 1-4\frac{r_{v_{\theta
}^{\max }}^{phys}}{R_{\max }^{phys}}\right) =0.8313
\end{equation*}
It results the Rossby radius $\rho _{g}=R_{\max }^{phys}/\left( \sqrt{2}%
L\right) \sim 212\,\left( km\right) $. The unit of vorticity is the Coriolis
parameter $\overline{\omega }=f_{0}=5\times 10^{-5}\;\left( s^{-1}\right) $
and we have the unit of velocity $\overline{v}=\rho _{g}\overline{\omega }%
=10.6\;\left( m/s\right) $ . Looking again to the results from the numerical
integration for $L=0.83$ we find the magnitude of the normalized tangential
velocity $v_{\theta }^{\max }=8.37$ (note that Eq.(\ref{vthmaxL}) gives a
similar value, $8.26$), which means that in physical units we have 
\begin{equation*}
v_{\theta \max }^{phys}\sim 88.6\;\left( m/s\right)
\end{equation*}
This gives a very high value for the maximum tangential wind, but the range
is still realistic. The maximum vorticity is $\left| \omega _{\max }\right|
=290.38\times f_{0}=0.0145\,\left( s^{-1}\right) $.

\subsection{Example 3}

We take for the third example the approximative value $R_{\max
}^{phys}/r_{v_{\theta }^{\max }}^{phys}\sim 8$ and a radius of maximum
extension of the hurricane of $R_{\max }^{phys}\sim 350\;\left( km\right) $.
(This is inspired by the picture taken by NASA on the hurricane Rita,
September 21, 2005). From this data it is obtained 
\begin{equation*}
L\sim \frac{1}{\sqrt{2}}\left( -2\right) \ln \left( 1-4\frac{r_{v_{\theta
}^{\max }}^{phys}}{R_{\max }^{phys}}\right) =0.9803
\end{equation*}
Then we can calculate the Rossby radius $\rho _{g}=R_{\max }^{phys}/\left( 
\sqrt{2}L\right) =252.47\;\left( km\right) $. Taking the Coriolis parameter $%
f_{0}=5\times 10^{-5}\;\left( s^{-1}\right) $ we have the physical unit of
velocity $\overline{v}=\rho _{g}\overline{\omega }=12.6\;\left( m/s\right) $
. The maximum tangential velocity for $L=0.98$ can be calculated using the
scaling formula Eq.(\ref{vthmaxL}), with the result $v_{\theta }^{\max
}=6.24 $. We will use however the exact result, provided by numerical
integration with GIANT of Eq.(\ref{eq}), $v_{\theta }^{\max }=6.15$ and this
leads to the physical velocity 
\begin{equation*}
v_{\theta \max }^{phys}=77.5\;\left( m/s\right)
\end{equation*}
The maximum vorticity is $\left| \omega _{\max }\right| =158.34\times
f_{0}=0.0079\,\left( s^{-1}\right) $.

\bigskip

These examples confirm the possibility of using the scaling relationships
Eq.(\ref{vthmaxL}) and Eq.(\ref{eyefit2}) to estimate the physical
charactersistics of the tropical cyclone. We note that the results of the
estimations can be significantly affected by the approximations on the
observational data, especially that of the ratio $r_{v_{\theta }^{\max
}}/R_{\max }$. This is because the magnitude of $L$ which is obtained from
Eq.(\ref{eyefit2}) using a reasonable input value for this ratio belongs to
a range where the variations with $L$ of all the characteristics of the
solution ($\psi $, $\omega $, $v_{\theta }$, $r_{v_{\theta }^{\max }}$) are
substantial. It is sufficient to look at the dependence of $v_{\theta
}^{\max }$ on $L$, Fig.\ref{fig_11}.

\section{Discussion}

At the origin of our approach it is the Charney-Hasegawa-Mima model, a
two-dimensional, nondissipative and purely fluid-dynamical (no thermal
process) model. Although is a simplified model it exhibits (via the
field-theoretical formulation) a compact analytic and algebraic structure,
self-duality, leading to Eq.(\ref{eq}). We have several arguments in favor
of the conclusion that Eq.(\ref{eq}) may represent the fluid
nonlinear-dynamic part of the atmospheric vortex. First, the profiles
obtained by solving Eq.(\ref{eq}) are similar to results already known (from
observations or numerical simulation) for the same quantities.

\begin{enumerate}
\item The profile of the the tangential velocity, our Fig.\ref{fig_5}, is
similar to typical tropical cyclone velocity profiles, as represented in
Fig.2 of Wang and Wu 2004. This is also confirmed by close similarity with
the Fig.1a from Reasor and Montgomery 2001 and with the profile given by the
Holland model or with the experimental observation (for Andrew hurricane,
Willoughby and Black 1996).

\item The fast decay to zero of the vorticity $\omega $ shown in our Fig.\ref%
{fig_2} is similar to what is shown in Fig.1a of Kossin and Schubert 2001;
ideally (without dissipation and friction) for mature hurricanes the maximum
of $\omega $ is on the center.

\item We note that in \ a series of reported numerical simulations, the
tendency of the fields is to evolve toward profiles that are very close to
those shown in our figures \ref{fig_2} and \ref{fig_5}. For example, the
Fig.7a and 7b of Kossin and Schubert 2001 show the evolution of the
vorticity and mean of the tangential velocity from initial profiles which
correspond to a narrow ring of vorticity to profiles that show clear
ressemblance with our figures \ref{fig_2} and \ref{fig_5}. The same striking
evolution to profiles similar to ours appears in Figs.7 a and b of the same
Reference. We have investigated whether a radially annular profile of
vorticity can be a solution of our equation (\ref{eq}). The result is
negative, which may explain why such an initial profile evolves to either a
set of vortices (vortex-crystal) or to a centrally peaked structure as in
Fig.\ref{fig_2}.
\end{enumerate}

Second, we obtain a good consistency between our quantitative results for an
atmospheric vortex (using approximative input information) and the values
measured or obtained in numerical simulations.

Finally, the quasi-solutions which appear to be an interesting feature of
this equation, are compatible with a series of previously known results: the
four vortices represented in our figure \ref{fig_10} are similar to the
Figure 4a from the work of Kossin and Schubert 2001. And the evolution to a
monopolar structure we obtain is similar to the same process reported in
this reference.

\bigskip

The numerical results have made possible to formulate several scaling
relationships connecting parameters of the atmospheric vortex. These are
simple formulas, intended for practical use, inevitably approximative. This
is because our purpose was to infer an analytical expression for a curve
which is determined numerically (or: is known in the form of a simple table
of values). Other expressions are possible and are worth to look for. A true
scaling law will become possible only as a result of an \emph{analytical }\
investigation of the properties of the equation. On the other hand, this
equation does not have a Backlund transform and is probable not integrable
by the Inverse Scattering Transform.

\bigskip

An important field of future investigation (analytical and numerical) is
represented by the \emph{quasi-solutions}, elements of a space of functions
that are very close of verifying the equation. Are they close to the
extremum of the action functional of the full field-theoretical model, for
example, are they metastable states? From the physical point of view it may
result that these states can be rendered more stable by processes that are
connected with what is missing from the Charney-Hasegawa-Mima model: third
dimension, viscosity, thermal processes. It is worth to examine the role of
the strongly localized quasi-solutions in models of tornadoes.

An interesting suggestion results from the massive series of radial
integrations: it appears that in the space of functions there are strings of
vortical quasi-solutions emanating from the exact solution and showing
increasing degree of departure from exactness (increase of the $error$
functional). If some physical factor will provide stability to these
quasi-solutions, then the system may slide along this string, with the
consequence that there is no pure stationarity but a continuous evolution to
stronger and stronger localisation of the vortex. This requires further
study.

Developing from the present one, a future self-consistent model will have to
include variation of the Rossby radius with the dynamical properties of the
vortex. Since this implies to consider that the coefficient of the
Chern-Simons part in the Lagrangian is a nonlinear function of the scalar
field, it is difficult to say if the self-duality will be maintained.

\bigskip

The investigation of this equation, and, most important, of the
field-theoretical model from which it is derived, are worth to be continued.

\textbf{Acknowledgments}. This work has been partly supported by the
Romanian Ministry of Education and Research and by the Japan Society for the
Promotion of Science. The hospitality of Professor S.-I. Itoh and of
Professor M. Yagi at the Kyushu University is gratefully acknowledged.

\end{document}